\documentclass[12pt]{article}
\pdfoutput=1
\usepackage[english]{babel}
\usepackage{graphicx}
    \usepackage[T1]{fontenc}
\usepackage{textcomp}
\usepackage{amsmath}
\usepackage{amssymb}
\usepackage{amsthm}
\usepackage{xcolor}
\usepackage{hyperref}
\usepackage{subcaption}
\usepackage{tikz}
\usepackage{setspace}
\usepackage{authblk}
\usepackage{natbib}

\usepackage[margin=1in]{geometry}
\defcitealias{Schorfheide2015}{SS15}

\title{Simulation smoothing for nowcasting with large mixed-frequency VARs\footnote{Corresponding author, e-mail: sebastian.ankargren@statistics.uu.se}}
\author{Sebastian Ankargren}
\author{Paulina Jon\'{e}us}
\affil{Department of Statistics, Uppsala University}

\begin{document}
\maketitle

\begin{abstract}
There is currently an increasing interest in large vector autoregressive (VAR) models. VARs are popular tools for macroeconomic forecasting and use of larger models has been demonstrated to often improve the forecasting ability compared to more traditional small-scale models. Mixed-frequency VARs deal with data sampled at different frequencies while remaining within the realms of VARs. Estimation of mixed-frequency VARs makes use of simulation smoothing, but using the standard procedure these models quickly become prohibitive in nowcasting situations as the size of the model grows. We propose two algorithms that alleviate the computational efficiency of the simulation smoothing algorithm. Our preferred choice is an adaptive algorithm, which augments the state vector as necessary to sample also monthly variables that are missing at the end of the sample. For large VARs, we find considerable improvements in speed using our adaptive algorithm. The algorithm therefore provides a crucial building block for bringing the mixed-frequency VARs to the high-dimensional regime. 
\end{abstract}

\section{Introduction}
Macroeconomic forecasting is a central activity for many organizations. Central banks, treasuries, and other government institutions as well as non-governmental organizations such as banks and corporations rely on forecasts in order to assess the future development of the local and global economy. Accurate forecasts is therefore a necessity for making informed decisions that minimize adversarial effects and maximize the utility of the decisions that are to be made. There is a vast array of forecasting methods that can be employed for macroeconomic data, but one of the more prominent and popular tools used in macroeconomic forecasting is the vector autoregressive (VAR) model. It is a multivariate time series model relating current values of a vector of variables to lags of all of the variables in the model. In contrast to structural models relying heavily on economic theory, the VAR is a relatively agnostic model that is easy to use. 


Traditional vector autoregressive models make use of data that is sampled at the same frequency. Most often, VAR models used for forecasting or structural analysis are estimated on data with exclusively monthly or quarterly variables. In many cases, the inclusion of gross domestic product (GDP) determines whether or not the model is estimated on a quarterly basis. Mixed-frequency methods, in general, are attempts that aim to avoid having to aggregate, say, monthly data to the quarterly frequency when at least one of the included variables is sampled quarterly. One advantage is that the loss of information induced by the aggregation can be avoided, and information can be incorporated into the forecasting model in a more timely manner. 
%
%
Among the more prominent methods are dynamic factor models as developed by \cite{Mariano2003,Mariano2010,Marcellino2016} and mixed-data sampling (MIDAS) as introduced by \cite{Ghysels2006} and \cite{Ghysels2007}. Related literature has also considered vector autoregressive moving average (VARMA) models in the mixed-frequency setting, see \cite{Zadrozny2016}, \cite{Anderson2016} and \cite{Deistler2017}. The relevance of the inclusion of an MA component has also been studied for MIDAS regressions by \cite{Foroni2019}. 

The MIDAS regression approach is generalized to a VAR setting with a Bayesian approach by \cite{Ghysels2016}, and \cite{McCracken2015} propose a high-dimensional variant of the MIDAS-VAR. \cite{Gotz2016} exploit the model by \cite{Ghysels2016} for testing Granger causality using mixed-frequency data. They consider both reduced rank regressions and a Bayesian approach in situations with a large number of high-frequency observations per low-frequency period.

The approach considered in this paper is the mixed-frequency VAR using a state-space formulation. The primary empirical purpose of the model that motivates our work is the use of a large monthly data set for forecasting GDP growth. We focus on the methodology developed by \cite{Schorfheide2015}, who use Gibbs sampling and simulation smoothing to estimate the model in the spirit of \cite{Carter1994}. Other developments of the mixed-frequency VAR include \cite{Eraker2015}, who instead of simulation smoothing employ a single-move sampler. \cite{Qian2016} develop an alternative sampling strategy based on linear transformations of the data. \cite{Cimadomo2016} employ a mixed-frequency VAR with time-varying parameters and stochastic volatility to cope with a change in frequency, and \cite{Gotz2018} extend the \cite{Schorfheide2015} model to include stochastic volatility and time-varying intercepts.

The focus of this paper is on developing the simulation smoother proposed by \cite{Schorfheide2015} in a way that enhances its performance in nowcasting situations. The merits of the mixed-frequency VARs in nowcasting and forecasting have been demonstrated by e.g. \cite{Schorfheide2015,Eraker2015,Ankargren2018,Gotz2018}. These papers frame the problem as a latent variable problem, where the low-frequency series have an underlying, high-frequency series which we do not observe. The model thus relates the observed outcome to the unobserved latent variable in a measurement equation, while the VAR itself is specified at the high frequency. 

One of the key advantages of the mixed-frequency VAR is that it is, in essence, a standard VAR with an additional block in the Markov Chain Monte Carlo (MCMC) sampler. This additional block samples the unobserved latent variable from the conditional posterior, given the observed data and the parameters, by means of simulation smoothing. Next, given the unobserved latent variable, the remaining steps needed to estimate the model are the usual steps drawing from the posterior of the parameters. The important difference is that the conditional posteriors of the parameters here condition upon the draw of the latent variable, which makes the problem standard. Because our focus in this paper is on the simulation smoother used for sampling the latent variables, we do not dwell any further on the other blocks of the MCMC sampler.

Our contribution is that we develop a simulation smoother based on the proposal by \cite{Schorfheide2015} that better deals with ragged edges in the (mixed-frequency) data, i.e. when the end of the sample is unbalanced. The motivation is that in real-time forecasting situations, the information set that the forecaster possesses can vary from day to day: new data points for the different series are not obtained at the same time, but on different days. In a stylized data set, the forecaster knows at the beginning of April the outcomes of financial variables for March, but the numbers for other variables have not yet been released. In addition, some variables are subject to larger publication delays. At the beginning of April, the forecaster may therefore not know the February outcomes for slow variables. We demonstrate that in such situations, use of our algorithm is a necessity when the dimension of the model is large. On the other hand, if the data is not unbalanced, our algorithm reduces to the \cite{Schorfheide2015} procedure. There is, in other words, no loss associated with using our procedure. Moreover, our algorithm is purely a computational improvement. The results from the different algorithms are therefore identical given the same random number generator.

For small-scale models, the simulation smoothing algorithm that \cite{Schorfheide2015} proposed works well. However, there is a growing literature on large VAR models and their usefulness for forecasting, see for example \cite{Banbura2010,Koop2013,Carriero2019}. When the dimension of the model increases---either by the number of variables or the lag length---the simulation smoother becomes prohibitive in estimating the models. The method we propose for dealing with the ragged edge nature of data is intuitive: we sample only those variables which we must sample. The issue with the \cite{Schorfheide2015} sampler is that, for the ragged edge part of the data, it wastes a substantial amount of computational resources on sampling what is already known, i.e. fully observed monthly variables. We circumvent this problem by developing an adaptive simulation smoother, which augments the state vector with the missing variables only. This allows us to improve the computational efficiency for large-dimensional models. Our method applies to any mixed-frequency VAR estimated using data with ragged edges that employs simulation smoothing.

The rest of the paper is organized as follows. Section \ref{sec:background} describes the mixed-frequency VAR and the approach suggested by \cite{Schorfheide2015} in more detail, Section \ref{sec:theory} describes our two new algorithms amd Section \ref{sec:computational} provides some computational experiments illustrating the gains of our algorithms. Section \ref{sec:conc} concludes.

\section{Mixed-frequency VARs}
\label{sec:background}
To model data observed at mixed frequencies we follow \cite{Schorfheide2015,Ankargren2018,Gotz2018} and deal with the problem by postulating a VAR model at the high frequency. That is, we start with the VAR($p$) given by
\begin{align}
x_t=\Pi_c+\Pi_1x_{t-1}+\cdots+\Pi_px_{t-p}+u_t, \quad u_t \sim N(0, \Sigma_t) \label{hf}
\end{align}
where $x_t$, $\Pi_c$ and $u_t$ are $n\times 1$ and $\Pi_j$ are $n\times n$. The vector $x_t$ can be decomposed into monthly and quarterly variables as $x_t=(x_{m,t}', x_{q,t}')'$ of dimensions $n_m\times 1$ and $n_q\times 1$, respectively, where $n=n_m+n_q$. The model can be formulated in terms of the companion form as
\begin{align}
z_t=\begin{pmatrix}\Pi_1 & \Pi_2 & \cdots & \Pi_p \\ \multicolumn{3}{c}{I_{n(p-1)}} & 0_{n(p-1)\times n}\end{pmatrix} z_{t-1}+\begin{pmatrix}\Pi_c \\ 0_{n(p-1)\times 1}\end{pmatrix}+\begin{pmatrix}u_t\\ 0_{n(p-1)\times 1}\end{pmatrix}
\end{align}
where $z_t=(x_t', \dots, x_{t-p+1}')'$. For future use, let also $\Pi=(\Pi_c, \Pi_1, \dots, \Pi_p)$.

The empirical problem motivating our work is that in reality we only observe
\begin{align}
y_t=S_t\Lambda z_t
\end{align}
where $S_t$ is a deterministic $n_t\times n$ selection matrix and $\Lambda$ an $n\times pn$ aggregation matrix. For ease of exposition, we assume that the aggregation employed includes at most $p-1$ previous values.

The posterior distribution of interest is $p(\Pi, \Sigma, X|Y)$. The key feature of \cite{Schorfheide2015} is the use of the forward-filtering and backwards-smoothing algorithm \citep{Carter1994,FruhwirthSchnatter1994} in a Gibbs sampler. Such an approach utilizes the fact that $(\Pi, \Sigma)$ is conditionally independent of $Y$ given $X$. Hence, the basic Gibbs sampling strategy for the mixed-frequency VAR is to sample iteratively from:
\begin{align}
&p(\Pi, \Sigma|X)\\
&p(X|\Pi, \Sigma, Y).
\end{align}
Note that conditional on $X$, the first step is standard in the literature. Intuitively, the full method can be understood from the perspective of missing values. The $p(X|\Pi, \Sigma, Y)$ step produces a draw of the underlying monthly series of all variables observed quarterly. This draw is produced under the restriction that when aggregated according to the linear aggregation defined by $\Lambda$, the aggregate is equal to the observed outcomes. However, conditional on a particular draw of $X$, draws of $\Pi$ and $\Sigma$ can be produced as if this were the real data, leading to standard procedures.

The main computational obstacle in an off-the-shelf implementation of a mixed-frequency model employing the companion form formulation is to draw from the posterior of the underlying variable, $p(X|\Pi, \Sigma, Y)$. Two ways of circumventing the high dimensionality are to use the computationally convenient reformulation of the model suggested by \cite{Schorfheide2015} or the method proposed by \cite{Qian2016}. The companion form presented in the beginning of the section should generally be avoided in computations because of the extraneous computational burden it induces. We review next the procedure developed by \cite{Schorfheide2015}, which largely avoids the companion form in the computations. Because of our frequent referral to their procedure, we will refer to it as \citetalias{Schorfheide2015}.

The companion form expresses the model in such a way that all of the endogenous variables in the VAR enter the state equation of the state-space model. Furthermore, lagged values of the variables are also included, which yields a state vector of dimension $np\times 1$. In empirical applications dealing with monthly and quarterly data, it is usually the case that the set of monthly variables is fully observed up through a time point $T_b$. 

The \citetalias{Schorfheide2015} method exploits the balanced nature of the monthly part of the sample for $t=1, \dots, T_b$. Intuitively, the primary purpose of employing a state-space model formulation and its associated simulation smoother is that it provides a tractable way of imputing the missing observations of the quarterly variables through a data augmentation approach common in Bayesian statistics (see \citealp{Tanner1987}). The simulation smoothing sampling step consists of sampling from $p(X|\Pi, \Sigma, Y)$, but since 
\begin{align}
X=\begin{pmatrix}X_{m, 1:T_b} & X_{q, 1:T_b}\\
X_{m, T_b+1:T} & X_{q, T_b+1:T}\end{pmatrix}
\end{align}
it is evident that including the observed monthly variables in the state vector means that we are also drawing from $p(X_{m, 1:T_b}|\Pi, \Sigma, Y)$ --- a degenerate distribution with all of its mass on $Y_{m, 1:T_b}$. To alleviate the computational cost, \citetalias{Schorfheide2015} therefore formulate the model differently for $t=1, \dots, T_b$, where the $X_{m, 1:T_b}$ is moved out of the state equation and treated as exogenous and the dimension of the state vector is thereby reduced from $np\times 1$ to $n_q(p+1)\times 1$. The new formulation of the model does not alter the underlying statistical model, but is purely computational. Throughout, we will refer to the following as the \emph{compact} form:

\begin{equation}
\begin{aligned}
\begin{pmatrix}y_{m,t}\\y_{q,t}\end{pmatrix}&=\begin{pmatrix}0_{n_m\times n_q} & \Pi_{mq}\\ \multicolumn{2}{c}{S_{q,t}\Lambda_{qs}}\end{pmatrix}\begin{pmatrix}x_{q, t} \\ z_{q,t-1}\end{pmatrix}+\begin{pmatrix}\Pi_{mm}&\Pi_{cm}\\\multicolumn{2}{c}{0_{n_q\times (pn_m+1)}}\end{pmatrix} \begin{pmatrix}y_{m,t-p:t-1}\\ 1\end{pmatrix}+\begin{pmatrix}W_t^{(1, \bullet)}\\0_{n_q\times n}\end{pmatrix}\begin{pmatrix}e_{m,t}\\e_{q,t}\end{pmatrix}\\
\begin{pmatrix}x_{q, t} \\ z_{q,t-1}\end{pmatrix}&=\begin{pmatrix}\Pi_{qq} & 0_{n_q\times n_q} \\I_{pn_q} & 0_{pn_q\times n_q}\end{pmatrix}\begin{pmatrix}z_{q, t-1}\\x_{q,t-p-1}\end{pmatrix}+\begin{pmatrix}\Pi_{qm}&\Pi_{cq}\\\multicolumn{2}{c}{0_{pn_q\times (pn_m+1)}} \end{pmatrix}\begin{pmatrix}y_{m,t-p:t-1}\\ 1\end{pmatrix}+\begin{pmatrix}W_t^{(2, \bullet)}\\0_{pn_q\times n}\end{pmatrix}\begin{pmatrix}e_{m,t}\\e_{q,t}\end{pmatrix},\label{eq:compact}
\end{aligned}
\end{equation}
where $e_t=\begin{pmatrix}e_{m,t}' & e_{q,t}'\end{pmatrix}'$ is independently and identically distributed as $N(0, I_n)$ and 
\begin{align}
\Sigma_t=W_tW_t', \quad W_t=\begin{pmatrix}W_t^{(1, \bullet)}\\W_t^{(2, \bullet)}\end{pmatrix}
\end{align}
and the partitioning in $W_t$ consists of blocks of size $n_m\times n$ and $n_q\times n$. Furthermore, the matrices $\Pi_{ij}$ for $i,j\in \{m, q\}$ represent submatrices of $\Pi$ consisting of parameters relating the set of $j$ variables to the $i$ variable; i.e. $\Pi_{mq}$ contains the parameters pertaining to the lagged effects of the quarterly variables in the equations for the monthly variables. In the preceding description of the model we explicitly allow for a time-varying error covariance matrix, but assume constant regression parameters. Relaxing the assumption of constant regression parameters, as \cite{Cimadomo2016,Gotz2018} have done for mixed-frequency VARs, can easily be done within the algorithms we develop.

To simplify the description of the filtering and smoothing algorithm, we write the model in more general notation as
\begin{align}
y_t&=Z_t \alpha_t+c_t+G_te_t\\
\alpha_t&=T_t\alpha_{t-1}+d_t+H_te_t.
\end{align}

The full \citetalias{Schorfheide2015} procedure for producing a draw from the conditional posterior uses the \cite{Durbin2002} simulation smoothing technique, which consists of the steps:\footnote{The constants in the model need to be excluded either in the simulation step, or in the filtering/smoothing steps, see \cite{Jarocinski2015}.}
\begin{enumerate}
\item[] Generate a pseudo sample $y_t^+$ and $\alpha_t^+$ based on the recursions in \eqref{hf}. For the difference $y_t^*=y_t-y_t^+$:
\item Filter for $t=1, \dots, T_b$ using the compact form
\item Move to the companion form and filter for $t=T_b+1, \dots, T$
\item Smooth for $t=T, \dots, T_b+1$ using the companion form
\item Move back to the compact form and smooth for $t=T_b, \dots, 1$
\item[] Compute the draw as $x_t=\hat{x}_t^*+x_t^+$ where $\hat{x}_t^*$ is the smoothed series from step 4.
\end{enumerate}

Because $n_q(p+1)\ll np$ in most applications, the compact procedure typically offers a substantial improvement in speed. 

The key parts of the simulation smoothing algorithm are the filtering and smoothing steps. In the following when discussing the simulation smoothing step, we will for simplicity discuss filtering and smoothing on their own, but note that they should be interpreted as the main building blocks for the full simulation smoothing algorithm. The filtering recursions are, for $t=1,\dots, T_b$:
\begin{equation}
\begin{alignedat}{3}
v_t&= y_t - Z_t a_t-c_t,&\quad M_t&=P_tZ_t'+H_tG_t', &\quad F_t&=Z_tM_t+G_t(G_t+Z_tH_t)'\\
K_t&=T_tM_tF_t^{-1}, &\quad  L_{t+1}&=T_{t+1}-K_tZ_t, &\quad N_{t+1}&=P_tL_{t+1}'-H_tG_t'K_t'\\
a_{t|t}&=a_t+M_t F_t^{-1}v_t, &\quad P_{t|t}&=P_t-M_tF_t^{-1}M_t'\\
a_{t+1}&=T_ta_{t|t}+d_t, &\quad P_{t+1} &=T_tP_{t|t}T_t'+H_tH_t'.
\end{alignedat}\label{eq:filt}
\end{equation}
Having transitioned back to the compact form, the algorithm is completed by the following smoothing recursions for $t=T_b, \dots, 1$:
\begin{equation}
\begin{aligned}
a_{t|T}&=a_{t|t}+N_{t+1}r_t\\
r_{t-1}&=L_{t+1}'r_t+Z_tF_t^{-1}v_t.
\end{aligned}\label{eq:smooth}
\end{equation}

The details for the transitions between the compact and companion forms can be found in \ref{sec:trans}.

\section{Handling mixed frequencies in large-dimensional VARs}
\label{sec:theory}

Unsurprisingly, if $n_q$ is fixed and $n$ grows an increasingly large share of the total computational time of the \citetalias{Schorfheide2015}  procedure can be attributed to the companion form, even if the companion form is employed for only one or two time points. The main bottleneck is the computation of the variance of the one-step ahead prediction. The variance is computed by multiplying three square matrices of dimensions $n(p+1)\times n(p+1)$, an operation which is prohibitive even if it needs to performed only once or twice if, say, $n=100$ and $p=4$. 

In the next two subsections, we describe two ways to increase the computational efficiency of filtering and smoothing for the unbalanced part of the sample. The first is \emph{blocked filtering}, in which the structure of the system matrices in the state-space model is exploited to reduce large matrix multiplications to a larger number of smaller block operations. The second approach is \emph{adaptive filtering}, which adaptively augments the state vector with the missing monthly variables and by doing so avoids the use of a full-blown companion form altogether. It should be emphasized that the two alternative procedures that we propose produce draws which, given the same seed, are numerically identical to draws made by the algorithm described in the previous section. 

For use in the following, we formulate the state-space model for the companion form as:\footnote{When the companion form is used for the unbalanced part of the data following use of the compact form, it is formulated with $p+1$ lags to aid in the transition between the model formulations.}
\begin{align}
y_t&=Z_t\alpha_t\\
\alpha_t&=T_t\alpha_{t-1}+H_t\epsilon_t,
\end{align}
where
\begin{align}
Z_t=S_t\begin{pmatrix}I_{n_m} & 0_{n_m\times (np+n_q)}\\ 0_{n_m\times n_m} & \Lambda_q \end{pmatrix}, &\quad \alpha_t = \begin{pmatrix}x_t' & \cdots & x_{t-p-1}'\end{pmatrix}'\label{eq:cmpn_obs}\\
T_t=\begin{pmatrix}\Pi_1 & \cdots & \Pi_p & 0_{n\times n} \\
\multicolumn{2}{c}{I_{np}} & 0_{n\times n} & 0_{n\times n}\end{pmatrix}, &\quad H_t=\begin{pmatrix}\Sigma_t^{1/2} & 0_{n\times np} \\ 0_{np\times n} & 0_{np\times np}\end{pmatrix}, \quad \epsilon_t=\begin{pmatrix}e_t \\ 0_{np\times1}\end{pmatrix}.\label{eq:cmpn_state}
\end{align}

For simplicity, we abstract here from including the intercept in the following, but inclusion thereof is straightforward.

\subsection{Blocked filtering}

While the bottleneck of the Kalman filter lies in the multiplication of three potentially large matrices, improvements can be made throughout the entire filtering algorithm by exploiting the fact that many of the system matrices in general are not dense, but have inherent structures---including blocks of zeros and identity matrices. \cite{Strid2009} successfully used the same idea to reduce the computational burden associated with a large-scale dynamic stochastic general equilibrium (DSGE) model. Inspired by their work for DSGE models, it is clear from Equations \eqref{eq:cmpn_obs}--\eqref{eq:cmpn_state} that there is a large degree of structure in the recursions for the companion form. Hence, it is likely subject to a speed improvement if the structure is efficiently utilized. In addition to what \eqref{eq:cmpn_obs}--\eqref{eq:cmpn_state} display, the matrix $\Lambda_q$ has a sparse structure where at most $n_qp$ columns (out of $np+n_q$) contain non-zero elements, the precise number depending on the specific aggregation scheme used. Let $p_q$ denote the number of lags included in the aggregation scheme. If the intra-quarterly average is used for aggregation, then $p_q=3$ and $3n_q$ columns contain non-zero values with a maximum of one non-zero value per column. Thus, when the system matrices become larger, substantial gains in computational efficiency can be obtained by simply exploiting these structures. 

Let $\Lambda_{qq}$ be the $n_q\times n_qp_q$ submatrix of $\Lambda_q$ obtained by deleting zero-columns in $\Lambda_q$, i.e. columns corresponding to monthly variables in $\alpha_t$ and additional lags not included in the aggregation. Let also $a_t$ (and $a_{t|t}$) be partitioned into $p+1$ blocks of $n$ values, i.e. $a_t=(a_t^{1'}, a_t^{2'}, \dots, a_t^{p+1'})'$. We present next the filtering algorithm for the companion form where the use of the structure is more explicit. In doing so, we let square brackets indicate that the enclosed quantity is computed before it is used to facilitate multiple uses efficiently. 

First, we can utilize the sparsity of $Z_t$ when computing $F_t$:
\begin{align}
F_t&=\begin{pmatrix}P_t^{mm} & [P_t^{mq}\Lambda_{qq}']\\
[P_t^{mq}\Lambda_{qq}']' & \Lambda_{qq} P_t^{qq} \Lambda_{qq}'\end{pmatrix}.
\end{align}
Here, $P_t^{mm}$ refers to the upper left $n_m\times n_m$ block of $P_t$ containing the variances of the one-step ahead predictions for the monthly variables. Similarly, $P_t^{qq}$ denotes the $p_qn_q\times p_qn_q$ submatrix of $P_t$ formed by extracting the elements corresponding to the first $p_q$ lags of all quarterly variables, and $P_t^{mq}$ is the $n_m\times p_qn_q$ submatrix of covariances between $x_{m,t}$ and $x_{q,t-i}$ for $i=0, \dots, p_q-1$. 

Next, compute $P_t^{1:p, m}(F_t^{-1})_{m, \bullet}$, where $P_t^{1:p, m}$ is the $np\times n_m$ submatrix retrieved from the upper left part of $P_t$, and $(F_t^{-1})_{m,\bullet}$ is the first $n_m$ rows of $F_t^{-1}$. Analogously, we compute also $P_t^{1:p, q}(F_t^{-1})_{q, \bullet}$ where $P_t^{1:p, q}$ is the $np\times n_q$ matrix obtained from taking the first $np$ rows of $P_t$ and the $n_q$ columns corresponding to $x_{q,t}$, and $(F_t^{-1})_{q, \bullet}$ is the bottom $n_q$ rows of $F_t^{-1}$. We can then construct:
\begin{align}
 K_t&=\begin{pmatrix}\Pi [P_t^{1:p, m}(F_t^{-1})_{m,\bullet}] & \Pi [P_t^{1:p, q}(F_t^{-1})_{q,\bullet}]\\
 [P_t^{1:p, m}(F_t^{-1})_{m,\bullet}] &  [P_t^{1:p, q}(F_t^{-1})_{q,\bullet}]\end{pmatrix}.
 \end{align}
Let the left block of columns in $K_t$ be denoted by $K_t^m$ and the right by $K_t^q$ and compute
 \begin{align}
 L_t&=\begin{pmatrix}\Pi_1 & \cdots & \Pi_p & 0_{n\times n} \\
\multicolumn{2}{c}{I_{np}} & 0_{np\times n} & 0_{np\times n}\end{pmatrix}-\begin{pmatrix}K_t^m & [K_t^q\Lambda_{q}]\end{pmatrix}
\end{align}
and
\begin{align}
M_t&=\begin{pmatrix}P_t^{1:(p+1), m} & P_t^{1:(p+1), q}\Lambda_{qq}'\end{pmatrix}.
\end{align}
The state prediction is
\begin{align}
a_{t+1}&=\begin{pmatrix}\Pi a_{t|t}^{1:p}\\ a_{t|t}^{2:(p+1)}\end{pmatrix}.\end{align}
Finally, a sizable improvement can be obtained for the variance of the state prediction by first computing $\Pi P_{t|t}^{1:p, 1:p}$ and then constructing
\begin{align}
P_{t+1} &=\begin{pmatrix}[\Pi P_{t|t}^{1:p, 1:p}]\Pi'+\Sigma_t & [\Pi P_{t|t}^{1:p, 1:p}]\\ [\Pi P_{t|t}^{1:p, 1:p}]'& P_{t|t}^{1:p, 1:p}\end{pmatrix}\label{eq:block_pt1}.
\end{align}
There are two main points to make from the recursions for blocked filtering. First, there are several instances where the same quantity appears in multiple places. It is therefore unnecessary to compute it over and over. Instead, efficiency can be gained by computing it once and copying the object to the appropriate locations. Second, because of the large number of zeros as well as the identity matrices in the system matrices, several computations can be avoided altogether in favor of subsetting the appropriate matrices. The two approaches can be used in conjunction in some cases. For example, to compute $K_tZ_t$ one would first select the first $n_m$ columns of $K_t$. To compute the second part of $K_tZ_t$, i.e. $K_t^q\Lambda_q$, one can compute $K_t^q\Lambda_{qq}$ and subtract the column vectors of the obtained matrix from the corresponding columns of $T_t$; because $\Lambda_q$ typically contains a large number of zero columns, so will $K_t^q\Lambda_q$ and these need not be computed. The same type of computation can also be carried out for $v_t$, which involves $Z_ta_t$ for which copying of pre-computed data and multiplication for a subset improves efficiency. 

The smoothing step is cheap in comparison with the filtering part, but is nevertheless subject to a modest improvement through:
\begin{align}
r_{t-1}&=L_{t+1}'r_t+\begin{pmatrix}[F_t^{-1}v_t]_{m, \bullet} \\\Lambda_q'[F_t^{-1}v_t]_{q, \bullet} \\ 0_{np\times 1}\end{pmatrix},
\end{align}
where $[F_t^{-1}v_t]_{m, \bullet}$ and $[F_t^{-1}v_t]_{q, \bullet}$ are the monthly and quarterly, respectively, blocks of rows of $F_t^{-1}v_t$, which is computed during filtering.

\subsection{Adaptive filtering}
The advantage of blocked filtering is its ability to improve upon the existing \citetalias{Schorfheide2015} compact filtering procedure with relatively little effort. However, the downside is that it improves on the problem without attempting to fix it.

\begin{figure}
\centering
\input{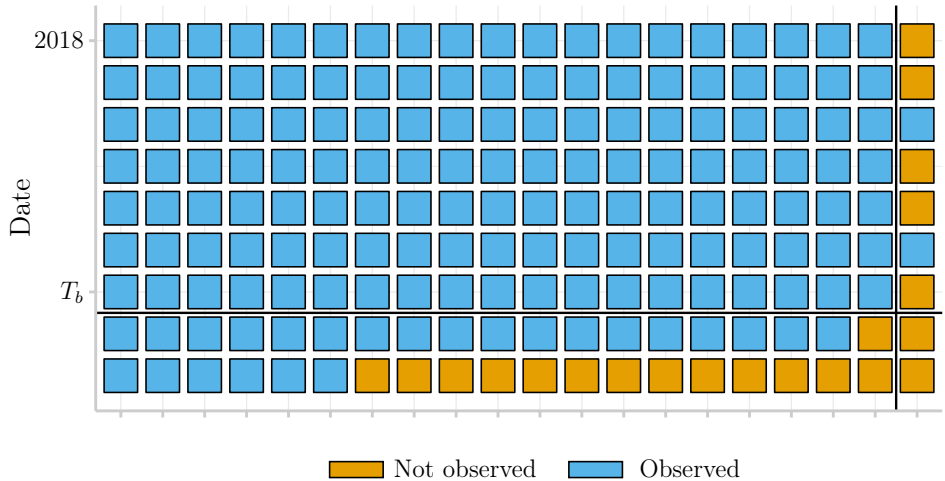}
\caption{Observational pattern for U.S. data. The color of each cell indicates if the variable is observed at the given time point. The right-most variable is a quarterly variable and the remaining are monthly indicators.}
\label{fig:obs}
\end{figure}

Figure \ref{fig:obs} offers an illustrative visual understanding of the compact filtering procedure: for $t\leq T_b$, the monthly block is fully observed, but for $t>T_b$ some of the monthly variables are missing. In this particular data set, based on the January vintage of the FRED database \citep{McCracken2016}, at $t=T_b+1$ only a single monthly indicator is missing. Nevertheless, in the compact procedure a move to the companion form is still triggered. Using the companion form, the 18 monthly variables which are observed at $t=T_b+1$ will have their true values imputed with themselves. For $t=T_b+2$, a larger number of monthly variables are unobserved. Yet, use of the full companion form would still need to carry out additional computations for the sole purpose of imputing six variables with their already-known outcomes.

Motivated by the large degree of unnecessary computations performed in the companion form for larger models, we therefore develop an adaptive procedure. Intuitively, the adaptive filtering procedure is simple and appealing: for every $t=T_b+1, \dots, T$, augment the state vector with the monthly variables missing at that time period. In relation to Figure \ref{fig:obs}, the procedure consists of adding the single missing monthly variable to the state vector at $t=T_b+1$ and at $t=T_b+2$ include all but the first six monthly variables in the state vector.

While the adaptive procedure is intuitive and simple in principle, it warrants some additional notation to be introduced. Let $\mathcal{U}_t$ denote the set of indexes of the monthly set of variables which are unobserved at time $t$. Correspondingly, let $\mathcal{O}_t$ denote the set of indexes of the observed set of monthly variables at time $t$. For $t=1, \dots, T_b$, we have that $y_{m,t}=y_{\mathcal{O}_t, t}$ is of dimension $n_m\times 1$. For $t=T_b+1, \dots, T$, it still holds that $y_{m,t}= y_{\mathcal{O}_t, t}$, but now the dimension is $|\mathcal{O}_t|<n_m$. The notation $y_{\mathcal{O}_t, t-\tau}$ is taken to mean the observed values at time $t-\tau$ of the monthly variables observed at time $t$. Similarly, $x_{\mathcal{U}_t, t-\tau}$ refers to the latent process at time $t-\tau$ for the set of unobserved monthly variables that are unobserved at time $t$.

Equipped with this new set of notation, we can formulate the state-space model as
\begin{equation}
\begin{aligned}
\begin{pmatrix}y_{m,t}\\y_{q,t}\end{pmatrix}&=Z_t\begin{pmatrix}x_{\mathcal{U}_t,t}\\x_{q,t} \\\vdots \\ x_{\mathcal{U}_t, t-p}\\x_{q,t-p}\end{pmatrix}+C_t \begin{pmatrix}y_{\mathcal{O}_{t-1},t-p:t-1}\\ 1\end{pmatrix}+G_te_t\\
\begin{pmatrix}x_{\mathcal{U}_t,t}\\x_{q,t} \\\vdots \\ x_{\mathcal{U}_t, t-p}\\x_{q,t-p}\end{pmatrix}&=T_t\begin{pmatrix}x_{\mathcal{U}_{t-1},t-1}\\x_{q,t} \\\vdots \\ x_{\mathcal{U}_{t-1}, t-p-1}\\x_{q,t-p-1}\end{pmatrix}+D_t\begin{pmatrix}y_{\mathcal{O}_{t-1},t-p:t-1}\\ 1\end{pmatrix}+H_te_t.
\end{aligned}
\end{equation}

The preceding model formulation is valid for any $t=1, \dots, T$, with no particular distinction between the balanced and unbalanced part. However, because $\mathcal{U}_t=\varnothing$ for $t=1, \dots, T_b$, the model is mostly constant and equal to the compact formulation. The \citetalias{Schorfheide2015} compact-companion procedure fits into the above formulation by specifying $\mathcal{U}_t=\varnothing$ and $\mathcal{O}_{t-1}=\{1, \dots, n_m\}$ for $t=1, \dots, T_b$, and $\mathcal{U}_t=\{1, \dots, n_m\}$ and $\mathcal{O}_{t-1}=\varnothing$ for $t=T_b+1, \dots, T$. One particular feature is that if $\mathcal{U}_t\neq \mathcal{U}_{t-1}$, $T_t$ will not be a square matrix. It is this time-varying size and shape that automatically lets the model augment new variables with missing observations to the state vector.

Let $J_t$ be a $(|\mathcal{U}_t|+n_q)\times(|\mathcal{U}_{t-1}|+n_q)$ selection matrix with the property that 
\begin{align}
J_t'\begin{pmatrix}x_{\mathcal{U}_t, t}\\x_{q, t}\end{pmatrix}=\begin{pmatrix}x_{\mathcal{U}_{t-1}, t}\\x_{q, t}\end{pmatrix}.
\end{align}
We can then construct the $T_t$ matrix as:
\begin{align}
T_t=\begin{pmatrix} \Pi_1^{\mathcal{U}_t, \mathcal{U}_{t-1}} & \Pi_1^{\mathcal{U}_t, q} & \cdots & \Pi_p^{\mathcal{U}_t, \mathcal{U}_{t-1}} & \Pi_p^{\mathcal{U}_t, q} & 0_{|\mathcal{U}_t|\times |\mathcal{U}_{t-1}|}\\
\multicolumn{5}{c}{I_p\otimes J_t} & 0_{n_q\times |\mathcal{U}_{t-1}|}
\end{pmatrix},
\end{align}
where $\Pi_j^{\mathcal{U}_t,\mathcal{U}_{t-1}}$ refers to the elements in the $\mathcal{U}_{t-1}$ columns and $\mathcal{U}_{t}$ rows of $\Pi_j$. In practice, there is no need to compute the Kronecker product as the bottom-left block of $T_t$ can more efficiently be filled programmatically. Depending on the observational pattern of the data, it may be worthwhile to use a block approach when dealing with multiplications involving $T_t$ because of the sparse structure of $I_p\otimes J_t$. However, if $n_q$ is low---our interest lies primarily in $n_q=1$---and if the unbalanced part contains many observations, there is likely little to gain.

The selection matrix $J_t$ is the $|\mathcal{U}_t|+n_q$ identity matrix with columns corresponding to $\mathcal{U}_t\bigcap\mathcal{O}_{t-1}$ deleted, i.e. columns for variables which were observed at $t-1$ but are unobserved at $t$. The matrix formed by collecting the deleted columns constitutes the orthogonal complement of $J_t$, denoted by $J_{\perp, t}$. The $J_{\perp, t}$ matrix is used in constructing $D_t$ to ensure that variables observed at $t-1$ but unobserved at $t$ get their lag identities---i.e. the bottom equations of the state equation---correct:
\begin{align}
D_t=\begin{pmatrix}\Pi_1^{\mathcal{U}_t, \mathcal{O}_{t-1}} & \cdots & \Pi_p^{\mathcal{U}_t, \mathcal{O}_{t-1}}\\
\Pi_1^{q, \mathcal{O}_{t-1}} & \cdots & \Pi_1^{q, \mathcal{O}_{t-1}}\\
\multicolumn{3}{c}{I_p\otimes J_{\perp, t}}
\end{pmatrix}.
\end{align}

For the observation equation, let $I_{-m, |\mathcal{U}_t|+n_q}$ be the $|\mathcal{U}_t|+n_q$ identity matrix with the first $|\mathcal{U}_t|$ rows deleted, so that $I_{-m, |\mathcal{U}_t|+n_q}\begin{pmatrix}x_{\mathcal{U}_t, t}\\x_{q,t}\end{pmatrix}=x_{q,t}$. Then we can construct the $Z_t$ system matrix as:
\begin{align}
Z_t=\begin{pmatrix}
0_{n_m\times(|\mathcal{U}_t|+n_q)} & \Pi_1^{\mathcal{O}_t,\mathcal{U}_t} &\Pi_1^{\mathcal{O}_t,q}  &\cdots &\Pi_p^{\mathcal{O}_t,\mathcal{U}_t} &\Pi_p^{\mathcal{O}_t,q}\\
\multicolumn{6}{c}{S_{q,t}\Lambda_{qq} [I_{p+1}\otimes I_{-m, |\mathcal{U}_t|+n_q}]}
\end{pmatrix}
\end{align}

\begin{align}
C_t=\begin{pmatrix}
\Pi_1^{\mathcal{O}_t, \mathcal{O}_{t-1}} & \cdots & \Pi_p^{\mathcal{O}_t, \mathcal{O}_{t-1}}\\
\multicolumn{3}{c}{S_{q,t}0_{n_q\times p|\mathcal{O}_{t-1}|}}
\end{pmatrix}
\end{align}

The only pieces missing for concluding the description of the adaptive procedure are the $H_t$ and $G_t$ matrices, which consist of:
\begin{align}
G_t&=\begin{pmatrix}
(\Sigma_t^{1/2})^{\mathcal{O}_t,\bullet}\\
S_{q,t}0_{n_q\times n}
\end{pmatrix}\\
H_t&=\begin{pmatrix}
(\Sigma_t^{1/2})^{\mathcal{U}_t,\bullet} \\
(\Sigma_t^{1/2})^{q, \bullet}\\
0_{p(|\mathcal{U}_t|+n_q)\times n}
\end{pmatrix},
\end{align}
where the superscripts refer to all columns and the subsets of rows corresponding to: the observed monthly variables ($\mathcal{O}_t$), the unobserved monthly variables ($\mathcal{U}_t$) or the quarterly variables ($q$).

It should be noted that while the system matrices can be characterized using zero, identity and selection matrices, it is usually inefficient to use these forms explicitly. Faster handling of the matrices and the operations involved can be achieved by subsetting and copying previously computed objects appropriately. 

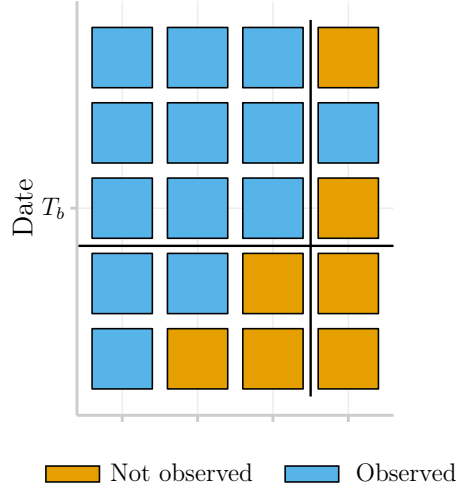
\begin{figure}
\centering
\begin{tikzpicture}[x=1pt,y=1pt]
\definecolor{fillColor}{RGB}{255,255,255}
\path[use as bounding box,fill=fillColor,fill opacity=0.00] (0,0) rectangle (216.81,198.74);
\begin{scope}
\path[clip] ( 60.06, 36.95) rectangle (179.79,193.74);
\definecolor{drawColor}{gray}{0.92}

\path[draw=drawColor,line width= 0.5pt,line join=round] ( 60.06,115.35) --
	(179.79,115.35);

\path[draw=drawColor,line width= 0.5pt,line join=round] ( 77.16, 36.95) --
	( 77.16,193.74);

\path[draw=drawColor,line width= 0.5pt,line join=round] (105.67, 36.95) --
	(105.67,193.74);

\path[draw=drawColor,line width= 0.5pt,line join=round] (134.17, 36.95) --
	(134.17,193.74);

\path[draw=drawColor,line width= 0.5pt,line join=round] (162.68, 36.95) --
	(162.68,193.74);
\definecolor{drawColor}{RGB}{0,0,0}
\definecolor{fillColor}{RGB}{86,180,233}

\path[draw=drawColor,line width= 0.6pt,line join=round,fill=fillColor] (122.77,160.96) rectangle (145.58,183.77);

\path[draw=drawColor,line width= 0.6pt,line join=round,fill=fillColor] (122.77,132.45) rectangle (145.58,155.26);

\path[draw=drawColor,line width= 0.6pt,line join=round,fill=fillColor] (122.77,103.95) rectangle (145.58,126.75);
\definecolor{fillColor}{RGB}{230,159,0}

\path[draw=drawColor,line width= 0.6pt,line join=round,fill=fillColor] (122.77, 75.44) rectangle (145.58, 98.24);

\path[draw=drawColor,line width= 0.6pt,line join=round,fill=fillColor] (122.77, 46.93) rectangle (145.58, 69.74);
\definecolor{fillColor}{RGB}{86,180,233}

\path[draw=drawColor,line width= 0.6pt,line join=round,fill=fillColor] ( 94.26,160.96) rectangle (117.07,183.77);

\path[draw=drawColor,line width= 0.6pt,line join=round,fill=fillColor] ( 94.26,132.45) rectangle (117.07,155.26);

\path[draw=drawColor,line width= 0.6pt,line join=round,fill=fillColor] ( 94.26,103.95) rectangle (117.07,126.75);

\path[draw=drawColor,line width= 0.6pt,line join=round,fill=fillColor] ( 94.26, 75.44) rectangle (117.07, 98.24);
\definecolor{fillColor}{RGB}{230,159,0}

\path[draw=drawColor,line width= 0.6pt,line join=round,fill=fillColor] ( 94.26, 46.93) rectangle (117.07, 69.74);
\definecolor{fillColor}{RGB}{86,180,233}

\path[draw=drawColor,line width= 0.6pt,line join=round,fill=fillColor] ( 65.76,160.96) rectangle ( 88.56,183.77);

\path[draw=drawColor,line width= 0.6pt,line join=round,fill=fillColor] ( 65.76,132.45) rectangle ( 88.56,155.26);

\path[draw=drawColor,line width= 0.6pt,line join=round,fill=fillColor] ( 65.76,103.95) rectangle ( 88.56,126.75);

\path[draw=drawColor,line width= 0.6pt,line join=round,fill=fillColor] ( 65.76, 75.44) rectangle ( 88.56, 98.24);

\path[draw=drawColor,line width= 0.6pt,line join=round,fill=fillColor] ( 65.76, 46.93) rectangle ( 88.56, 69.74);
\definecolor{fillColor}{RGB}{230,159,0}

\path[draw=drawColor,line width= 0.6pt,line join=round,fill=fillColor] (151.28,160.96) rectangle (174.08,183.77);
\definecolor{fillColor}{RGB}{86,180,233}

\path[draw=drawColor,line width= 0.6pt,line join=round,fill=fillColor] (151.28,132.45) rectangle (174.08,155.26);
\definecolor{fillColor}{RGB}{230,159,0}

\path[draw=drawColor,line width= 0.6pt,line join=round,fill=fillColor] (151.28,103.95) rectangle (174.08,126.75);

\path[draw=drawColor,line width= 0.6pt,line join=round,fill=fillColor] (151.28, 75.44) rectangle (174.08, 98.24);

\path[draw=drawColor,line width= 0.6pt,line join=round,fill=fillColor] (151.28, 46.93) rectangle (174.08, 69.74);

\path[draw=drawColor,line width= 0.9pt,line join=round] (148.43,186.62) --
	(148.43, 44.08);

\path[draw=drawColor,line width= 0.9pt,line join=round] ( 60.06,101.09) -- (179.79,101.09);
\end{scope}
\begin{scope}
\path[clip] (  0.00,  0.00) rectangle (216.81,198.74);
\definecolor{drawColor}{gray}{0.80}

\path[draw=drawColor,line width= 1.1pt,line join=round] ( 60.06, 36.95) --
	( 60.06,193.74);
\end{scope}
\begin{scope}
\path[clip] (  0.00,  0.00) rectangle (216.81,198.74);
\definecolor{drawColor}{RGB}{0,0,0}

\node[text=drawColor,anchor=base east,inner sep=0pt, outer sep=0pt, scale=  0.80] at ( 55.56,112.59) {$T_b$};
\end{scope}
\begin{scope}
\path[clip] (  0.00,  0.00) rectangle (216.81,198.74);
\definecolor{drawColor}{gray}{0.80}

\path[draw=drawColor,line width= 1.1pt,line join=round] ( 57.56,115.35) --
	( 60.06,115.35);
\end{scope}
\begin{scope}
\path[clip] (  0.00,  0.00) rectangle (216.81,198.74);
\definecolor{drawColor}{gray}{0.80}

\path[draw=drawColor,line width= 1.1pt,line join=round] ( 60.06, 36.95) --
	(179.79, 36.95);
\end{scope}
\begin{scope}
\path[clip] (  0.00,  0.00) rectangle (216.81,198.74);
\definecolor{drawColor}{gray}{0.80}

\path[draw=drawColor,line width= 1.1pt,line join=round] ( 77.16, 34.45) --
	( 77.16, 36.95);

\path[draw=drawColor,line width= 1.1pt,line join=round] (105.67, 34.45) --
	(105.67, 36.95);

\path[draw=drawColor,line width= 1.1pt,line join=round] (134.17, 34.45) --
	(134.17, 36.95);

\path[draw=drawColor,line width= 1.1pt,line join=round] (162.68, 34.45) --
	(162.68, 36.95);
\end{scope}
\begin{scope}
\path[clip] (  0.00,  0.00) rectangle (216.81,198.74);
\definecolor{drawColor}{RGB}{0,0,0}

\node[text=drawColor,rotate= 90.00,anchor=base,inner sep=0pt, outer sep=0pt, scale=  0.90] at ( 43.22,115.35) {Date};
\end{scope}
\begin{scope}
\path[clip] (  0.00,  0.00) rectangle (216.81,198.74);
\definecolor{drawColor}{RGB}{255,255,255}
\definecolor{fillColor}{RGB}{255,255,255}

\path[draw=drawColor,line width= 0.4pt,line join=round,line cap=round,fill=fillColor] ( 30.20,  5.00) rectangle (209.64, 24.45);
\end{scope}
\begin{scope}
\path[clip] (  0.00,  0.00) rectangle (216.81,198.74);
\definecolor{drawColor}{RGB}{0,0,0}
\definecolor{fillColor}{RGB}{230,159,0}

\path[draw=drawColor,line width= 0.6pt,line cap=round,fill=fillColor] ( 48.29, 11.71) rectangle ( 68.55, 17.74);
\end{scope}
\begin{scope}
\path[clip] (  0.00,  0.00) rectangle (216.81,198.74);
\definecolor{drawColor}{RGB}{0,0,0}
\definecolor{fillColor}{RGB}{86,180,233}

\path[draw=drawColor,line width= 0.6pt,line cap=round,fill=fillColor] (138.79, 11.71) rectangle (159.05, 17.74);
\end{scope}
\begin{scope}
\path[clip] (  0.00,  0.00) rectangle (216.81,198.74);
\definecolor{drawColor}{RGB}{0,0,0}

\node[text=drawColor,anchor=base east,inner sep=0pt, outer sep=0pt, scale=  0.80] at (126.70, 11.97) {Not observed};
\end{scope}
\begin{scope}
\path[clip] (  0.00,  0.00) rectangle (216.81,198.74);
\definecolor{drawColor}{RGB}{0,0,0}

\node[text=drawColor,anchor=base east,inner sep=0pt, outer sep=0pt, scale=  0.80] at (203.64, 11.97) {Observed};
\end{scope}
\end{tikzpicture}
\caption{Observational pattern for the example discussed in the text. The color of each cell indicates if the variable is observed at the given time point. The right-most variable is a quarterly variable and the remaining are monthly indicators.}
\label{fig:obs2}
\end{figure}

The procedure itself is simple and straightforward, but the description of the system matrices is somewhat involved notationally. To illustrate how the model adapts to the structure of observed data, let us demonstrate it using a simple example. Figure \ref{fig:obs2} displays the observational pattern for our hypothetical data. The system matrices are not displayed here for space considerations, but can be found in full in Appendix \ref{app:example}.

As before, the compact form is used for the balanced part of the sample as $\mathcal{U}_t=\varnothing$ and $\mathcal{O}_{t-1}=\{1, 2, 3\}$ for $t=1, \dots, T_b$. We use the short form $y_{1:j, t-1:t-3}$ to denote the vector $y_{1:j, t-1:t-3}=(y_{1, t-1}, \dots, y_{j, t-1}, y_{1, t-2}, \dots, y_{j, t-3})'$. For the current example, the observation and state equations for a model with three lags are:
\begin{equation}
\begin{aligned}
\begin{pmatrix}y_{1,t}\\y_{2,t}\\ y_{3,t} \\ y_{q,t}\end{pmatrix}&=Z_t\begin{pmatrix}x_{q, t} \\ x_{q,t-1}\\ x_{q,t-2} \\ x_{q,t-3}\end{pmatrix}+C_t y_{1:3, t-1:t-3}+G_t e_t\\
\begin{pmatrix}x_{q, t} \\ x_{q,t-1}\\ x_{q,t-2} \\ x_{q,t-3}\end{pmatrix}&=T_t\begin{pmatrix}x_{q, t-1} \\ x_{q,t-2}\\ x_{q,t-3} \\ x_{q,t-4}\end{pmatrix}+D_t y_{1:3, t-1:t-3}+H_te_t.
\end{aligned}
\end{equation}

At $t=T_b+1$, the third monthly variable is missing so that $\mathcal{U}_{T_b+1}=\{3\}$ and $\mathcal{O}_{T_b}=\{1, 2, 3\}$. The third monthly variable is thus added to the state equation:
\begin{equation}
\begin{aligned}
\begin{pmatrix}y_{1,T_b+1}\\y_{2,T_b+1}\end{pmatrix}&=Z_{T_b+1}\begin{pmatrix}x_{q, {T_b+1}} \\ x_{q,{T_b}}\\ x_{q,{T_b}-1} \\ x_{q,{T_b}-2}\end{pmatrix}+C_{T_b+1} y_{1:3, T_b:T_b-2}+G_{T_b+1}e_{T_b+1}\\
\begin{pmatrix}x_{3, T_b+1} \\ x_{q, {T_b+1}} \\ x_{3, T_b} \\ x_{q,{T_b}}\\ x_{3, T_b-1} \\ x_{q,{T_b}-1} \\ x_{3, T_b-2} \\x_{q,{T_b}-2}\end{pmatrix}&=T_{T_b+1}\begin{pmatrix}x_{q, {T_b+1}} \\ x_{q,{T_b}}\\ x_{q,{T_b}-1} \\ x_{q,{T_b}-2}\end{pmatrix}+D_{T_b+1} y_{1:3, T_b:T_b-2}+H_{T_b+1}e_{T_b+1}.
\end{aligned}
\end{equation}

At the end of the sample ($t=T_b+2=T)$, we obtain $\mathcal{U}_{T_b+2}=\{2, 3\}$ and $\mathcal{O}_{T_b+1}=\{1, 2\}$. The second monthly variable is now added to the state equation:
\begin{equation}
\begin{aligned}
y_{1,T_b+2}&=Z_{T_b+2}\begin{pmatrix}x_{3, T_b+1} \\ x_{q, {T_b+1}} \\ x_{3, T_b} \\ x_{q,{T_b}}\\ x_{3, T_b-1} \\ x_{q,{T_b}-1} \\ x_{3, T_b-2} \\x_{q,{T_b}-2}\end{pmatrix}+C_{T_b+2} \begin{pmatrix}y_{1,T_b:T_b-2}\\ y_{2,T_b:T_b-2} \end{pmatrix}+G_{T_b+2}e_{T_b+2}\\
\begin{pmatrix}x_{2, T_b+1} \\x_{3, T_b+1} \\ x_{q, {T_b+1}} \\ x_{2, T_b} \\ x_{3, T_b} \\ x_{q,{T_b}}\\ x_{2, T_b-1} \\ x_{3, T_b-1} \\ x_{q,{T_b}-1} \\ x_{2, T_b-2} \\ x_{3, T_b-2} \\x_{q,{T_b}-2}\end{pmatrix}&=T_{T_b+2}\begin{pmatrix}x_{3, T_b+1} \\ x_{q, {T_b+1}} \\ x_{3, T_b} \\ x_{q,{T_b}}\\ x_{3, T_b-1} \\ x_{q,{T_b}-1} \\ x_{3, T_b-2} \\x_{q,{T_b}-2}\end{pmatrix}+D_{T_b+2} \begin{pmatrix}y_{1,T_b:T_b-2}\\ y_{2,T_b:T_b-2} \end{pmatrix}+H_{T_b+2}e_{T_b+2}.
\end{aligned}
\end{equation}

In principle, the procedure thus simply amounts to continuing to perform the compact filtering procedure, but adjusting which variables are included as exogenous based on the observational pattern. The filtering and smoothing recursions are therefore the same as for the compact procedure in \eqref{eq:filt}--\eqref{eq:smooth}; what changes is instead the composition in the model and the construction of the system matrices.

By studying the recursions in more detail, the benefit of the adaptive procedure becomes clear. Take the main bottleneck---the computation of the variance of the one-step ahead prediction---as an example. Using the companion form with blocked filtering for the observational pattern in Figure \ref{fig:obs} with $p=3$ requires multiplying $\Pi \, (20\times 80)$, $P_{t|t}^{1:p, 1:p}\,(80\times80)$ and the transpose $\Pi'$ together. The number of scalar multiplications required to compute the full product is $20^3\times 80^3=4,096,000,000$.\footnote{The calculated number of scalar multiplications does not exploit the symmetric nature of the resulting matrix.}  Using the adaptive procedure, it is (for $P_{T_b+2}$) instead necessary to multiply $T_{T_b+2}\, (14\times 8)$, $P_{T_b+1|T_b+1}\,(8\times 8)$ and the transpose $T_{T_b+2}'$ together. The number of scalar multiplications for this latter operation to be executed is $14^3\times 8^3=1,404,928$---almost four orders of magnitude smaller. There is, in other words, large possible gains to be achieved from adaptively adjusting the state vector to include only the bare minimum. The relative gain depends on the number of observed variables in the unbalanced part and the length of the unbalanced part.

\section{Computational experiments}
\label{sec:computational}
In this section, we demonstrate the computational benefits of our proposed algorithms. In designing the setting for which we conduct the computational experiments, we aim to stay close to a realistic scenario. Our primary empirical interest lies in using variables sampled monthly in forecasting a single quarterly variable, the canonical example in mind being use of a large set of monthly indicators for forecasting GDP growth. In order to see also the effects of including more quarterly variables (cf. \citealp{Schorfheide2015,Gotz2018}), we also consider $n_q=3$. For the total number of variables, we consider a grid of $n=10, 20, \dots, 120$ to capture the computational gains in both small and large models. The sample size is $T=500$ and we let the number of monthly variables with missing values at $t=499, 500$ be: one for $n=10, \dots, 40$, two for $n=50, \dots, 80$ and three for $n=90, \dots, 120$. The number of fully observed monthly series is set to $0.3n$ (rounded up), with the remaining containing missingness at $t=500$ only. The composition of the ragged edge is roughly in line with the January 2019 vintage of the FRED-MD database \citep{McCracken2016}. We first let $p=6$ and vary $n$, and then consider the reverse situation. We implement the simulation smoothing routines in C++ using the Armadillo library for linear algebra \citep{Sanderson2016} via R and RcppArmadillo \citep{R2019,Eddelbuettel2014}.


\begin{figure}
\centering
        \input{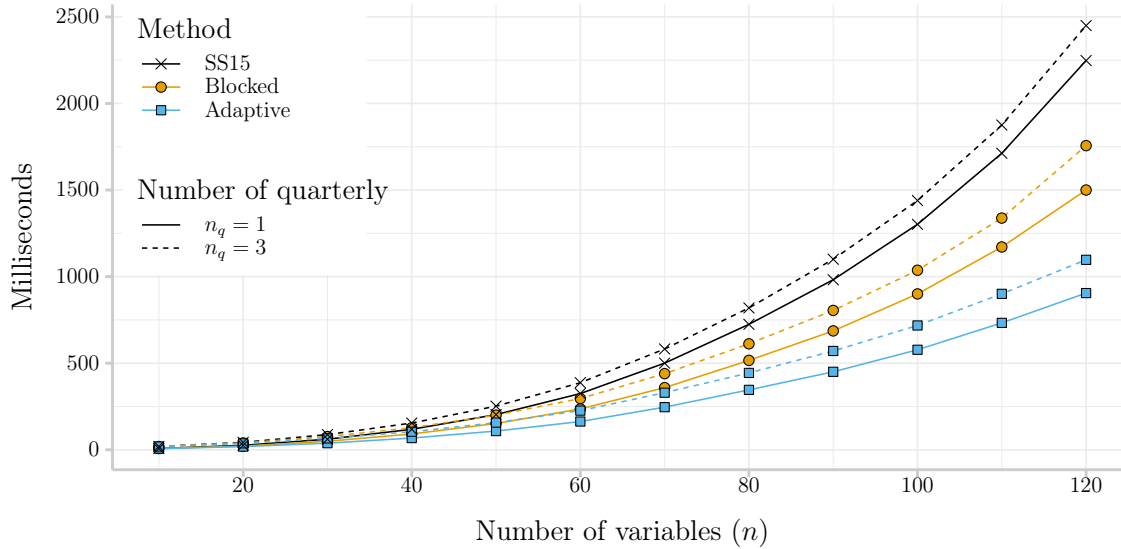}
        \caption{Computational cost as a function of $n$, milliseconds per iteration. The example uses $T=500$ and $T_b=498$. The number of monthly variables with missing observations at both $T_b+1$ and $T_b+2$ is $0.025n$ and the number of fully observed monthly series is $0.3n$ (both rounded up). The lag length is $p=6$.}
\label{fig:timing2}
\end{figure}

Figure \ref{fig:timing2} displays the computational cost per iteration. At the top in the figure, we find the standard \citetalias{Schorfheide2015} compact procedure followed by, in order, the blocked and adaptive approaches. It is evident from the figure that for low-dimensional models, there is no need to care about the computational efficiency as the cost per iteration is low to begin with. Nevertheless, when the dimension of the model is large the choice of algorithm starts to matter. There is also little relative difference between using one or three quarterly variables in the model when the model size is large, which means that in large models there is no need to restrict oneself from adding more quarterly variables to the model for computational reasons. For a model using a data set similar in size to \cite{Banbura2010} and \cite{Carriero2019}, use of the adaptive method yields 5,000 draws in approximately one hour, whereas the \citetalias{Schorfheide2015} procedure requires over 3 hours. While an hour is still needed, avoiding every bottleneck possible is highly desirable in a real-time forecasting situation where the model may be subject to frequent re-estimation.

\begin{figure}
\centering
\begin{subfigure}[b]{\textwidth}
        \centering
        \input{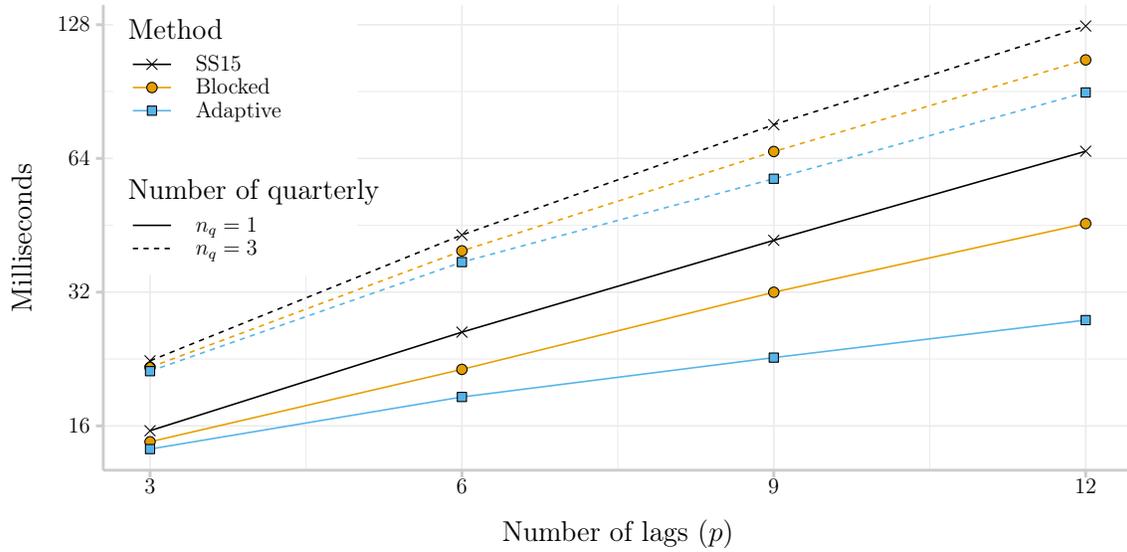}
        \caption{$n=20$}
    \end{subfigure}\\
\begin{subfigure}[b]{\textwidth}
        \centering
        \input{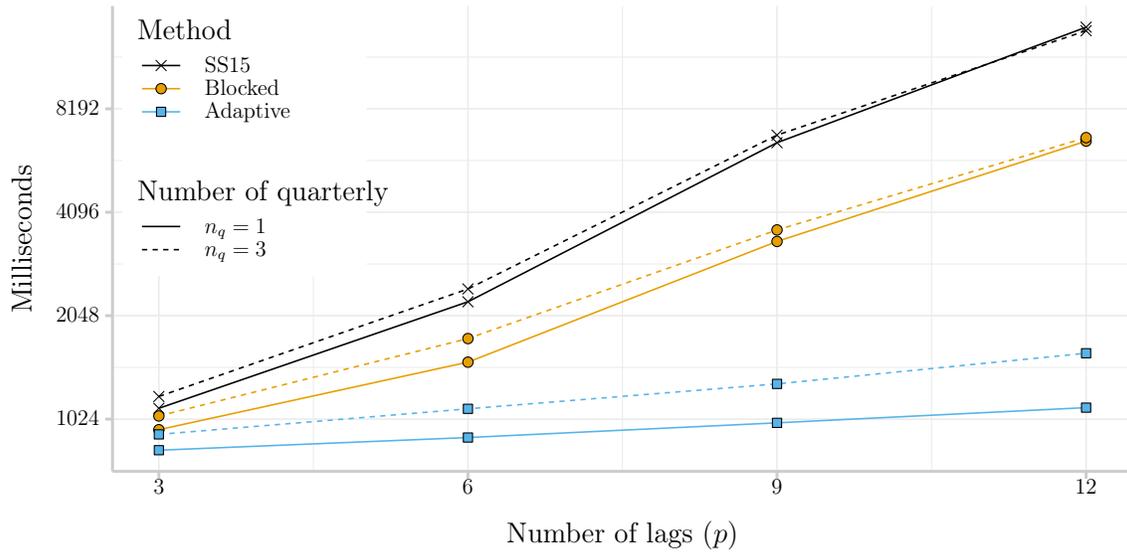}
        \caption{$n=120$}
    \end{subfigure}\\
\caption{Computational cost as a function of $p$, milliseconds per iteration. The example uses $T=500$ and $T_b=498$. The number of monthly variables with missing observations at both $T_b+1$ and $T_b+2$ is one ($n=20$) and four ($n=120$), and the number of fully observed monthly series is three ($n=10$) and 36 ($n=120$).}
\label{fig:timing3}
\end{figure}

Because the three methods all make use of the compact filtering procedure for the balanced part, the difference in computational cost is due to the different treatments of the unbalanced part. Thus, the figure clearly shows that using the companion form---even for two time periods only---can be costly, as substantial reductions can be obtained by handling it more carefully. In fact, using the adaptive procedure, dealing with the unbalanced part constitutes only a small part of the overall computational cost. Consequently, for further speed improvements one needs to focus on filtering and smoothing in the compact form. Further enhancements of the compact form may be possible depending on the remaining parts of the model, if e.g. there are structures imposed on the parameters that can be exploited.

Figure \ref{fig:timing3} displays the cost per iteration as a function of the number of lags $p$. The number of variables is $n=20$ (upper panel) and $n=120$ (bottom panel). Note that these two model sizes correspond to the models used by \cite{Banbura2010,Carriero2019}. In both of the aforementioned papers, the authors use $p=13$. The figure shows that as the number of lags increases, the added computational cost is lower using the adaptive procedure than if the blocked approach is used. Going from 3 to 12 lags roughly means an increase from 0.8 to 1.1 seconds using the adaptive procedure. In contrast, the cost of the \citetalias{Schorfheide2015} procedure increases from 1.1 seconds to 14.1 seconds. \cite{Carriero2019} report that producing 5,000 posterior draws in their large model takes about 5.5 hours; adding a quarterly variable to their model would increase the overall computational time to approximately 7 hours using the adaptive procedure. If instead \citetalias{Schorfheide2015} is used, estimating the model would take approximately 24 hours.

\begin{figure}
\centering
\input{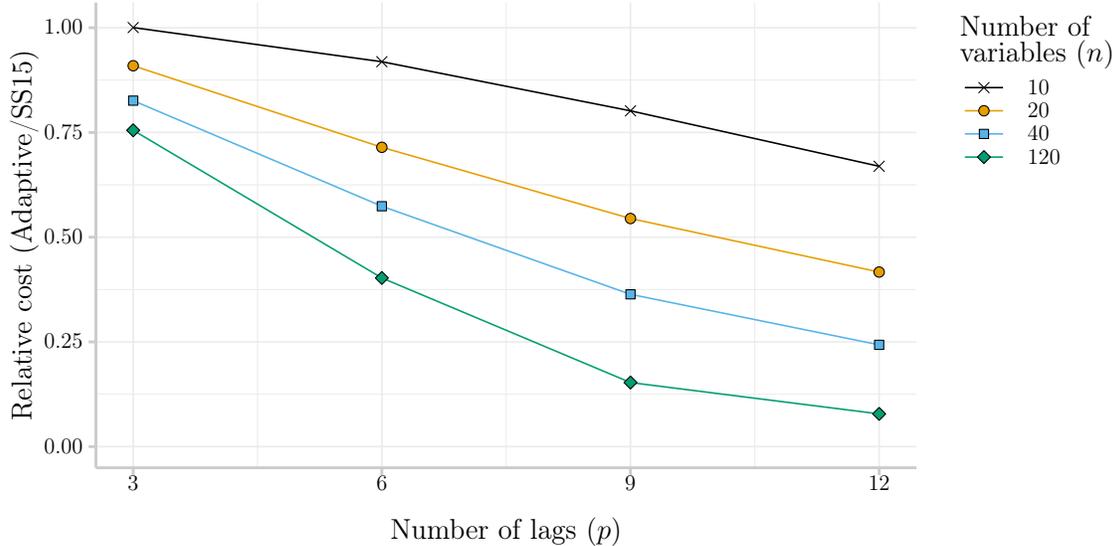}
\caption{Relative cost of the adaptive algorithm. The example uses $T=500$, $T_b=498$ and $n_q = 1$. The number of monthly variables with missing observations at both $T_b+1$ and $T_b+2$ is one ($n=10,20,40$) and three ($n=120$), and the number of fully observed monthly variables is three ($n=10$), six ($n=20$), 12 ($n=40$) and 36 ($n=120$).}
\label{fig:timing4}
\end{figure}

To more clearly see where gains of the adaptive algorithm are larger, Figure \ref{fig:timing4} presents the relative cost of the adaptive algorithm relative to the \citetalias{Schorfheide2015} algorithm. It more explicitly details the pattern that previous figures have documented: the adaptive algorithm scales better with $p$, and it is primarily for large-dimensional models that the adaptive algorithm excels and provides substantial improvements. The precise level of the gains also depends on the observational pattern in the data. While the \citetalias{Schorfheide2015} and the blocked procedures are unaffected by the specific observational pattern (given $T_b$ and $T$), the adaptive procedure is more apt the more variables are observed. Consequently, situations in which the data set includes a large number of financial variables that are observed immediately are especially suitable for the adaptive algorithm. The adaptive algorithm therefore makes use of large-dimensional VARs for nowcasting in the presence of mixed-frequency data feasible.

\section{Conclusion}
\label{sec:conc}
The increasing interest in large vector autoregressive models means that existing methods and algorithms are put to the test as large models are substantially more computationally demanding. In order to facilitate estimation of large-dimensional VARs with mixed-frequency data for real-time nowcasting we have developed methods that cope better with the large dimensionality. Our preferred algorithm is adaptive and augments the state vector in the simulation smoother at each time point as necessary, thereby avoiding to sample series we already know. The adaptive algorithm exploits the ragged edges of the data to ameliorate estimation when the data set is unbalanced. The algorithm consistently improves upon the \cite{Schorfheide2015} simulation smoother, with the largest gain obtained when the number of variables and the number of lags are high. Given that recent influential papers such as \cite{Banbura2010} and \cite{Carriero2019} estimate models with the number of variables being in the range 125--131 while the number of lags is set to 13, the situation in which we find the largest improvements is clearly of high interest in the literature. Our algorithm can therefore be used as an efficient building block in incorporating additional variables sampled at lower frequencies in existing large-dimensional VARs.


\bibliographystyle{apalikedoi}
\bibliography{all_refs}
\appendix

\section{Transitions in the \citetalias{Schorfheide2015} simulation smoother}
\label{sec:trans}
\paragraph{From compact to companion}
At $t=T_b$, we obtain $a_{T_b+1}$ and $P_{T_b+1}$ for $\alpha_t=(x_{q, t}', z_{q, t-1}')'$. If we instead use $\tilde{\alpha}_t=(z_t', x_{t-p}')'$, we see that these now represent
\begin{align}
\tilde{a}_{T_b|T_b}&=E\left(\begin{array}{c}x_{m, T_b} \\x_{q, T_b} \\x_{m, T_b-1} \\ x_{q, T_b-1} \\ \vdots \\x_{m, T_b-p} \\x_{q, t-p}\end{array}\bigg| y_{T_b}, y_{T_b-1}, \dots\right)=\left(\begin{array}{c}x_{m, T_b} \\ a_{T_b|T_b}^{(1)} \\x_{m, T_b-1} \\ a_{T_b|T_b}^{(2)} \\ \vdots \\x_{m, T_b-p} \\a_{T_b|T_b}^{(p+1)}\end{array}\right)
 \end{align} 
where $a_{t|t}=(a_{t|t}^{(1)'}, \dots, a_{t|t}^{(p)'})'$. Moreover,
\begin{align}
\tilde{P}_{T_b|T_b}&=V\left(\begin{array}{c}x_{m, T_b} \\x_{q, T_b} \\x_{m, T_b-1} \\ x_{q, T_b-1} \\ \vdots \\x_{m, T_b-p} \\x_{q, t-p}\end{array}\bigg| y_{T_b}, y_{T_b-1}, \dots\right)=\left(\begin{array}{cccccc}0 &0 &0 &0 & \cdots & 0 \\ 0&P_{T_b|T_b}^{(1,1)} & 0 & P_{T_b|T_b}^{(1,2)}&\cdots & P_{T_b|T_b}^{(1,p+1)}\\0&0&0&0&\cdots&0\\ 0&P_{T_b|T_b}^{(2,1)}&0 &P_{T_b|T_b}^{(2,2)} & \cdots & P_{T_b|T_b}^{(2,p+1)}\\ \vdots &\vdots & \vdots & \vdots & \ddots & \vdots\\ 0&P_{T_b|T_b}^{(p+1,1)}&0&P_{T_b|T_b}^{(p+1,2)} & \cdots& P_{T_b|T_b}^{(p+1,p+1)}\end{array}\right),
 \end{align} 
 where $P_{T_b|T_b}^{(i,j)}$ is the $n_q\times n_q$ $(i, j)$ block of $P_{T_b|T_b}$.
 
 Now, we can see that
 \begin{align}
\tilde{a}_{T_b+1}&= F_1(\Pi)\tilde{a}_{T_b|T_b}+F_c(\Pi)\\
\tilde{P}_{T_b+1}&= F_1(\Pi)\tilde{P}_{T_b|T_b}F_1(\Pi)'+\Omega(\Sigma_t).
 \end{align}

The companion-based filter is obtained by using $\tilde{a}_{T_b+1}$ and $\tilde{P}_{T_b+1}$ as starting values and then proceeding as usual. By smoothing, we obtain $\hat{\tilde{\alpha}}_s=E({\tilde{\alpha}}_{s}|y_T, y_{T-1}, \dots)$ for $s=T, \dots, T_b+1$. 

\paragraph{From companion to compact}
The final step is to move from the companion form to the compact representation. Doing so is accomplished by
\begin{align}
r_{T_b}=P_{T_b+1}^{-1}(\hat{\alpha}_{T_b+1}-a_{T_b+1})
\end{align}

Hence, to move from the companion form back to the compact representation, we extract $\hat{\alpha}_{T_b+1}=E(\alpha_{T_b+1}|y_T, y_{T-1}, \dots)$ from $\hat{\tilde{\alpha}}_{T_b+1}$. After that, we can smooth using \eqref{eq:smooth} initialized with $r_{T_b}$.

\section{Adaptive filtering: Example}
\label{app:example}
\paragraph{System matrices for $t=1, \dots, T_b$}
For $t=1, \dots, T_b$, monthly variables are fully observed and the quarterly variable is observed every third month. When $t$ corresponds to a month in which the quarterly variable is not observed, the fourth row of $Z_t$, $C_t$ and $G_t$ is deleted. We let $\Pi_{l}^{i,j}$ denote element $(i, j)$ of $\Pi_l$, and $w_t^{i,j}$ element $(i, j)$ of the lower-triangular Cholesky factor $W_t$. 
\begin{align}
Z_t = \begin{pmatrix}0 & \Pi_1^{1,4} & \Pi_2^{1,4} & \Pi_3^{1,4}\\
0 & \Pi_1^{2,4} & \Pi_2^{2,4} & \Pi_3^{2,4}\\
0 & \Pi_1^{3,4} & \Pi_2^{3,4} & \Pi_3^{3,4}\\
\frac{1}{3} & \frac{1}{3} & \frac{1}{3} & 0\end{pmatrix}, \quad C_t = 
\begin{pmatrix}\Pi_1^{1,1} & \Pi_2^{1,1} & \Pi_3^{1,1}&\Pi_1^{1,2} & \Pi_2^{1,2} & \Pi_3^{1,2}&\Pi_1^{1,3} & \Pi_2^{1,3} & \Pi_3^{1,3}\\
\Pi_1^{2,1} & \Pi_2^{2,1} & \Pi_3^{2,1}&\Pi_1^{2,2} & \Pi_2^{2,2} & \Pi_3^{2,2}&\Pi_1^{2,3} & \Pi_2^{2,3} & \Pi_3^{2,3}\\ 
\Pi_1^{3,1} & \Pi_2^{3,1} & \Pi_3^{3,1}&\Pi_1^{3,2} & \Pi_2^{3,2} & \Pi_3^{3,2}&\Pi_1^{3,3} & \Pi_2^{3,3} & \Pi_3^{3,3}\\ 
0&0&0&0&0&0&0&0&0\end{pmatrix}\\
G_t = \begin{pmatrix}w_t^{1,1}&0&0&0\\
w_t^{2,1} & w_t^{2,2} &0&0\\
w_t^{3,1} & w_t^{3,2} & w_t^{3,3} & 0
\\ 0&0&0 & 0\end{pmatrix}, \quad
T_t=\begin{pmatrix} \Pi_1^{4,4} & \Pi_2^{4,4} & \Pi_3^{4,4} & 0 \\
1 & 0 & 0 & 0\\
0 & 1 & 0 & 0\\
0 & 0 & 1 & 0 \end{pmatrix}\\
 D_t = \begin{pmatrix}\Pi_1^{4,1} & \Pi_2^{4,1} & \Pi_3^{4,1} & \Pi_1^{4,2} & \Pi_2^{4,2} & \Pi_3^{4,2} & \Pi_1^{4,3} & \Pi_2^{4,3} & \Pi_3^{4,3} \\ 0&0&0&0&0&0&0&0&0\\ 0&0&0&0&0&0&0&0&0\\ 0&0&0&0&0&0&0&0&0\end{pmatrix}, \quad H_t = \begin{pmatrix}w_t^{4,1} & w_t^{4,2} & w_t^{4,3} & w_t^{4,4}\\ 0 & 0 & 0 & 0 \\ 0 & 0 & 0 & 0 \\ 0 & 0 & 0 & 0 \end{pmatrix}
\end{align}

\paragraph{System matrices for $t= T_b+1$}
\begin{align}
Z_t = \begin{pmatrix}0 & \Pi_1^{1,4} & \Pi_2^{1,4} & \Pi_3^{1,4}\\
0 & \Pi_1^{2,4} & \Pi_2^{2,4} & \Pi_3^{2,4}\end{pmatrix}, \quad C_t = 
\begin{pmatrix}\Pi_1^{1,1} & \Pi_2^{1,1} & \Pi_3^{1,1}&\Pi_1^{1,2} & \Pi_2^{1,2} & \Pi_3^{1,2}&\Pi_1^{1,3} & \Pi_2^{1,3} & \Pi_3^{1,3}\\
\Pi_1^{2,1} & \Pi_2^{2,1} & \Pi_3^{2,1}&\Pi_1^{2,2} & \Pi_2^{2,2} & \Pi_3^{2,2}&\Pi_1^{2,3} & \Pi_2^{2,3} & \Pi_3^{2,3}\end{pmatrix}\\
G_t = \begin{pmatrix}w_t^{1,1}&0&0&0\\
w_t^{2,1} & w_t^{2,2} &0&0\end{pmatrix}, \quad
T_t=\begin{pmatrix} \Pi_1^{3,4} & \Pi_2^{3,4} & \Pi_3^{3,4} & 0 \\
\Pi_1^{4,4} & \Pi_2^{4,4} & \Pi_3^{4,4} & 0 \\
0 & 0 & 0 & 0 \\
1 & 0 & 0 & 0\\
0 & 0 & 0 & 0 \\
0 & 1 & 0 & 0\\
0 & 0 & 0 & 0 \\
0 & 0 & 1 & 0 \end{pmatrix}\\
 D_t = \begin{pmatrix}\Pi_1^{3,1} & \Pi_2^{3,1} & \Pi_3^{3,1} & \Pi_1^{3,2} & \Pi_2^{3,2} & \Pi_3^{3,2} & \Pi_1^{3,3} & \Pi_2^{3,3} & \Pi_3^{3,3} \\ \Pi_1^{4,1} & \Pi_2^{4,1} & \Pi_3^{4,1} & \Pi_1^{4,2} & \Pi_2^{4,2} & \Pi_3^{4,2} & \Pi_1^{4,3} & \Pi_2^{4,3} & \Pi_3^{4,3} \\0&0&0&0&0&0&1&0&0\\ 0&0&0&0&0&0&0&0&0\\ 0&0&0&0&0&0&0&1&0\\0&0&0&0&0&0&0&0&0\\ 0&0&0&0&0&0&0&0&1\\ 0&0&0&0&0&0&0&0&0\end{pmatrix}, \quad H_t = \begin{pmatrix}w_t^{3,1} & w_t^{3,2} & w_t^{3,3}  & 0 \\ w_t^{4,1} & w_t^{4,2} & w_t^{4,3} & w_t^{4,4}\\ 0 & 0 & 0 & 0 \\ 0 & 0 & 0 & 0\\ 0 & 0 & 0 & 0 \\ 0 & 0 & 0 & 0\\ 0 & 0 & 0 & 0 \\ 0 & 0 & 0 & 0 \end{pmatrix}
\end{align}

\paragraph{System matrices for $t= T_b+2$}
\begin{align}
Z_t = \begin{pmatrix}0 & 0 & \Pi_1^{1,3} & \Pi_1^{1,4} & \Pi_2^{1,3} & \Pi_2^{1,4} & \Pi_3^{1,3} &  \Pi_3^{1,4}\end{pmatrix}, \quad C_t = 
\begin{pmatrix}\Pi_1^{1,1} & \Pi_2^{1,1} & \Pi_3^{1,1} & \Pi_1^{1,2} & \Pi_2^{1,2} &\Pi_3^{1,2}  \end{pmatrix}\\
G_t = \begin{pmatrix}w_t^{1,1}&0&0&0\end{pmatrix}, \quad
T_t=\begin{pmatrix}\Pi_1^{2,3} & \Pi_1^{2,4} & \Pi_2^{2,3} &  \Pi_2^{2,4} & \Pi_3^{2, 3} & \Pi_3^{2,4} & 0 & 0 \\
\Pi_1^{3,3} & \Pi_1^{3,4} & \Pi_2^{3,3} &  \Pi_2^{3,4} & \Pi_3^{3, 3} & \Pi_3^{3,4} & 0 & 0 \\
\Pi_1^{4,3} & \Pi_1^{4,4} & \Pi_2^{4,3} &  \Pi_2^{4,4} & \Pi_3^{4, 3} & \Pi_3^{4,4} & 0 & 0 \\
0 & 0 & 0 & 0 &0 & 0 & 0 & 0 \\
1 & 0 & 0 & 0&0 & 0 & 0 & 0 \\
0 & 1 & 0 & 0&0 & 0 & 0 & 0 \\
0 & 0 & 0 & 0 &0 & 0 & 0 & 0 \\
0 & 0 & 1 & 0&0 & 0 & 0 & 0 \\
0 & 0 & 0 & 1&0 & 0 & 0 & 0 \\
0 & 0 & 0 & 0&0 & 0 & 0 & 0 \\
0 & 0 & 0 & 0&1 & 0 & 0 & 0 \\
0 & 0 & 0 & 0&0 & 1 & 0 & 0 \\ \end{pmatrix}\\
 D_t = \begin{pmatrix}\Pi_1^{2,1} & \Pi_2^{2,1} & \Pi_3^{2,1} & \Pi_1^{2,2} & \Pi_2^{2,2} & \Pi_3^{2,2} \\ \Pi_1^{3,1} & \Pi_2^{3,1} & \Pi_3^{3,1} & \Pi_1^{3,2} & \Pi_2^{3,2} & \Pi_3^{3,2} \\ \Pi_1^{4,1} & \Pi_2^{4,1} & \Pi_3^{4,1} & \Pi_1^{4,2} & \Pi_2^{4,2} & \Pi_3^{4,2}  \\0&0&0&1&0&0\\ 0&0&0&0&0&0\\ 0&0&0&0&0&0\\0&0&0&0&1&0\\ 0&0&0&0&0&0\\ 0&0&0&0&0&0\\0&0&0&0&0&1\\ 0&0&0&0&0&0\\ 0&0&0&0&0&0\end{pmatrix}, \quad H_t = \begin{pmatrix}
w_t^{2,1} & w_t^{2,2} & 0 &0 \\w_t^{3,1} & w_t^{3,2} & w_t^{3,3}  & 0 \\ w_t^{4,1} & w_t^{4,2} & w_t^{4,3} & w_t^{4,4}\\ 0 & 0 & 0 & 0 \\ 0 & 0 & 0 & 0\\ 0 & 0 & 0 & 0 \\ 0 & 0 & 0 & 0\\ 0 & 0 & 0 & 0 \\ 0 & 0 & 0 & 0\\ 0 & 0 & 0 & 0 \\ 0 & 0 & 0 & 0\\ 0 & 0 & 0 & 0 \end{pmatrix}
\end{align}

\end{document}